# Truck-Involved Crashes Injury Severity Analysis for Different Lighting Conditions on Rural and Urban Roadways

Majbah Uddin and Nathan Huynh*

University of South Carolina
Department of Civil and Environmental Engineering
300 Main St, Columbia, SC 29208, USA

*Corresponding Author Contact Information
Nathan Huynh
University of South Carolina
Department of Civil and Environmental Engineering
300 Main St, Columbia, SC 29208, USA
Telephone: (803) 777-8947
Fax: (803) 777-0670
Email: nathan.huynh@sc.edu

**Abstract**

This paper investigates factors affecting injury severity of crashes involving trucks for different lighting conditions on rural and urban roadways. It uses 2009–2013 Ohio crash data from the Highway Safety Information System. The explanatory factors include the occupant, vehicle, collision, roadway, temporal and environmental characteristics. Six separate mixed logit models were developed considering three lighting conditions (daylight, dark, and dark-lighted) on two area types (rural and urban). A series of log-likelihood ratio tests were conducted to validate that these six separate models by lighting conditions and area types are warranted. The model results suggest major differences in both the combination and the magnitude of impact of variables included in each model. Some variables were significant only in one lighting condition but not in other conditions. Similarly, some variables were found to be significant in one area type but not in other area type. These differences show that the different lighting conditions and area types do in fact have different contributing effects on injury severity in truck-involved crashes, further highlighting the importance of examining crashes based on lighting conditions on rural and urban roadways. Age and gender of occupant (who is the most severely injured in a crash), truck types, AADT, speed, and weather condition were found to be factors that have significantly different levels of impact on injury severity in truck-involved crashes.

# 1. Introduction

The trucking industry plays a vital element in freight logistics and economic well-being of a country. Furthermore, it has significant potential to increase economic productivity for shippers and carriers by ensuring timely and efficient flow of commodities. According to the Bureau of Transportation Statistics, trucks moved about 13,955 millions of tons of freight valued at more than $11,444 billion in the United States in 2013 (Bureau of Transportation Statistics, 2015). With the increase in truck volume, there is growing concerns related to traffic safety due to the magnitude of injury severity and economic impact of truck-involved crashes (Abramson, 2015; Lyman and Braver, 2003).

Trucks contribute to the large numbers of crashes, injuries, and fatalities because of their high volume on roadways, size, weight, and unique operating characteristics (e.g., longer braking distance) (Zhu and Srinivasan, 2011a). In 2013, 3,964 people were killed and another 95,000 were injured in crashes involving an estimated 342,000 trucks in the U.S. (NHTSA, 2015). According to the NHTSA report, truck drivers had the highest percentage of previously recorded crashes than drivers of any other type of vehicles. The cost associated with the truck-involved crashes can be substantial. Zaloshnja and Miller (2007) estimated that the average cost of a police-reported crash involving a truck is $91,112, based on 2005 dollars. Additionally, they estimated the average cost per fatality, non-fatality, and property damage only crashes to be $3,604,518, $195,258, and $15,114, respectively.

The safety and costs imposed on society by truck-involved crashes necessitates the need to better understand the underlying contributing factors so that counter measures can be developed to prevent or reduce such crashes. This study is focused on investigating the relationships between crash factors and crash injury severity, based on different area types (i.e., rural and urban) and lighting conditions which have not been studied previously. Past studies have identified significant differences between rural and urban crashes due to differing occupant, vehicle, roadway, and environmental characteristics (Islam et al., 2014; Khorashadi et al., 2005). Furthermore, several studies have indicated that roadway lighting conditions play a significant role in truck-involved crashes (Chang and Mannering, 1999; Chen and Chen, 2011; Duncan et al., 1998; Islam and Hernandez, 2013a, 2013b; Islam et al., 2014; Khattak et al., 2003;

Khorashadi et al., 2005; Lemp et al., 2011; Pahukula et al., 2015; Zhu and Srinivasan, 2011a). However, the limitation of these studies is that they capture the impact of lighting conditions via indicator variables representing different lighting conditions. The interaction between variables is complex which can vary significantly across different lighting conditions and area types. For instance, while the aggregate model may indicate that adverse weather increases injury severity of occupants, its effect may vary under different lighting conditions and area types. Adverse weather may contribute to severe injury at rural locations under dark condition, whereas in urban locations under daylight condition it may contribute to less severe injury. One possible reason for this is that poor visibility increases reaction time, and therefore potentially causing more severe injuries. To this end, this study aims to investigate the factors that influence injury severity of truck-involved crashes on both rural and urban roadways under different lighting conditions: daylight, dark (dark without street lights), and dark-lighted (dark with street lights).

In this study, mixed logit (random parameters logit) models are used to provide a better understanding of the interaction between crash factors found in the dataset and unobserved factors (i.e., unobserved heterogeneity). Mixed logit models are statistically superior to traditional fixed parameters logit models and they require less detailed crash-specific data than that of the fixed parameters models (Anastasopoulos and Mannering, 2011; Chen and Chen, 2011). To best of the authors' knowledge, this study is the first to examine the contributing factors to injury severity (major injury, minor injury and possible/no injury) by examining truck-involved crashes under different lighting conditions and area types.

## 2. Literature review

To date, there have been several research studies analyzing the injury severity of truck-involved crashes. Table 1 provides a summary of these studies considering data source, analysis region, factors included, inclusion of lighting variable, analysis methods used, and key research outcomes relevant to effects of lighting. Collectively, there are three commonalities among these studies. First, all of these studies considered injury severity as the dependent variable; some used injury severity of driver (Chang and Mannering, 1999; Chen and Chen, 2011; Dong et al., 2015; Islam and Hernandez, 2013b; Khorashadi et al., 2005; Pahukula et al., 2015) and some used injury severity of the most severely injured occupant involved in a crash (Duncan et al., 1998; Islam et al., 2014; Islam and Hernandez, 2013a; Lemp et al., 2011). Second, the existing body of work on truck safety explored the effects of lighting condition on injury severity of truck-involved crashes via the use of an independent indicator variable. However, this approach is limited since the interaction between lighting condition and injury severity levels is complex. Third, no study has investigated the effect of area types and lighting conditions exclusively on injury severity of truck-involved crashes. This study seeks to fill this knowledge gap by using mixed logit models to analyze truck-involved crashes. Specifically, the contributions of this study are: (i) to investigate the differences of effects of factors that contribute to injury severity in truck-involved crashes under three lighting conditions (daylight, dark, and dark-lighted) and two area types (rural and urban) and (ii) to demonstrate the necessity of using a disaggregate approach to analyze truck-involved crashes.

**Table 1**
Injury severity studies related to truck-involved crashes.

| Authors | Date source & region | Independent variables | Lighting variable | Model type | Key outcomes |
|---|---|---|---|---|---|
| Duncan et al. (1998) | Highway Safety Information System; North Carolina | Driver and roadway characteristics | ✓ | Ordered probit | Dark condition (DC) had a higher effect on injury risk to the occupants compared to other factors; dark-lighted condition (DLC) did not have any effect on injury severity |
| Chang and Mannering (1999) | Washington Department of Transportation; Washington | Driver, vehicle, roadway, temporal and environmental characteristics | | Nested logit | Crashes involving trucks had higher injury severity than those of non-truck-involved crashes; for truck-involved crashes the probability of injury or fatality was 50% higher if the crash occurred at night |
| Khattak et al. (2003) | Highway Safety Information System; North Carolina | Driver, vehicle, roadway, collision and environmental characteristics | ✓ | Binary probit, ordered probit | DC was found as one of the contributing factors to rollover and more severe injuries for truck crashes |
| Khorashadi et al. (2005) | California Department of Transportation; California | Driver, vehicle, roadway, temporal and environmental characteristics | ✓ | Multinomial logit | There was a 31% increase in the probability of severe/fatal injury for crashes under DC; the probability of severe/fatal injury for drivers in tractor-trailer combinations was 26% higher than that of single-unit trucks on rural roadways |
| Chen and Chen (2011) | Highway Safety Information System; Illinois | Driver, vehicle, roadway, collision, temporal and environmental characteristics | ✓ | Mixed logit | DC was found to be significant in explaining truck-involved crash injury severity in multi-vehicle truck crashes |
| Zhu and Srinivasan (2011a) | Large Truck Crash Causation Study; 17 U.S. states | Driver, collision and vehicle characteristics | ✓ | Ordered probit | The crashes occurred under DLC were found to be more severe than that of other lighting conditions for truck-involved crashes |
| Zhu and Srinivasan (2011b) | Large Truck Crash Causation Study; 17 U.S. states | Driver, collision and vehicle characteristics | | Heteroskedastic ordered probit | The use of illegal drugs, driving under influence, and inattention of the drivers were found to be significant factors that contribute to injury severity |
| Lemp et al. (2011) | Large Truck Crash Causation Study; 17 U.S. states | Driver, collision and vehicle characteristics | ✓ | Ordered probit, heteroskedastic ordered probit | There was an 8% increase in the probability of fatal injury for truck-involved crashes under DC; the same percentage of increase is found under DLC |
| Islam and Hernandez (2013a) | National Automotive Sampling System; U.S. | Occupant, vehicle, roadway, collision and environmental characteristics | ✓ | Ordered probit, random parameter ordered probit | For 76% of the crashes under DC, injuries sustained by the occupants were found to be less severe in truck-involved crashes |
| Islam and Hernandez (2013b) | Texas Peace Officers' Crash Reports; Texas | Driver, roadway, temporal and environmental characteristics | ✓ | Mixed logit | There was an 11% increase in the probability of fatal injury for truck-involved crashes under DC |
| Islam et al. (2014) | University of Alabama Center for Advanced Public Safety; Alabama | Driver, vehicle, roadway, collision, temporal and environmental characteristics | ✓ | Mixed logit | There was a 3% increase in the probability of major injury for urban multi-vehicle at-fault truck crashes under DC |
| Dong et al. (2015) | Tennessee Department of Transportation; Tennessee | Driver, vehicle, roadway and environmental characteristics | | Multinomial logit | Traffic that was lower in volume but with higher percentage of trucks contributed to severe/fatal injury; lighting condition was found not to be statistically significant |
| Pahukula et al. (2015) | Texas Peace Officers' Crash Reports; Texas | Driver, vehicle, roadway, collision, temporal and environmental characteristics | ✓ | Mixed logit | DC was found to be significant only for early morning (12 AM–4 AM) model; the probability of severe and minor injury is higher when lighting condition was dark |
| Osman et al. (2016) | Highway Safety Information System; Minnesota | Driver, vehicle, roadway, collision and temporal characteristics | | Multinomial logit, ordered logit, generalized ordered logit | Daytime crashes, no control of access, higher speed limits, and crashes on rural principal arterials were the most important factors that contribute to higher injury severity |

## 3. Methodology

There have been numerous studies that examined the relationship between crash factors and injury severity outcomes using discrete choice models, such as multinomial logit models, mixed logit models, and ordered logit/probit models (cf. Savolainen et al., 2011). This study uses mixed logit models for the reasons stated earlier. Specifically, its use is necessary to account for unobserved heterogeneity (unobserved factors) and its formulation does not have the independence of irrelevant alternatives (IIA) property of the standard multinomial logit model (Washington et al., 2003). Typically, crash injury severities are reported as discrete outcomes (e.g., major injury, minor injury, and possible/no injury). This ordered nature has led researchers to use the ordered logit/probit models (e.g., Abdel-Aty, 2003; Islam and Hernandez, 2013a; Zhu and Srinivasan, 2011a). However, the standard ordered models restrict the influence of explanatory variables on injury severity. That is, they either decrease the highest injury severity level and increase the lowest, or decrease the lowest injury severity level and increase the highest (Khorashadi et al., 2005; Kim et al., 2013). It should be noted that advanced versions of the ordered models such as the generalized ordered logit model and the partial proportional odds model can relax the above assumption (Savolainen et al., 2011).

Following the methodology presented in Milton et al. (2008), the relationship between the injury severity variable and the explanatory variables is expressed as follows.

$$Y_{in} = \beta_i X_{in} + \epsilon_{in} \tag{1}$$

where $Y_{in}$ is the variable representing injury severity level $i$ ($i \in I$ denotes injury severity levels, i.e., major injury, minor injury, and possible/no injury) of an individual $n$, $X_{in}$ is the injury severity explanatory variables/factors, $\beta_i$ is the parameter to be estimated for each injury severity level $i$, and $\epsilon_{in}$ is the error term to capture the effects of the unobserved characteristics. If the error term is independently and identically distributed with generalized extreme value distribution, then the resulting model is a multinomial logit model with the following choice probability.

$$P_n(i) = \frac{\exp[\beta_i X_{in}]}{\sum_{i \in I} \exp[\beta_i X_{in}]} \tag{2}$$

where $P_n(i)$ is the probability of injury severity level $i$ for individual $n$.

To capture the effects of unobserved heterogeneity due to randomness associated with some of the factors necessary to understand injury sustained by the occupants, Eq. (2) is extended to the following mixed logit model formulation (Train, 2009).

$$P_n(i|\phi) = \int \frac{\exp[\beta_i X_{in}]}{\sum_{i \in I} \exp[\beta_i X_{in}]} f(\beta_i|\phi) d\beta_i \tag{3}$$

where $P_n(i|\phi)$ is the probability of injury severity level $i$ conditional on $f(\beta_i|\phi)$, $f(\beta_i|\phi)$ is the density function of $\beta_i$ and $\phi$ is the parameter vector with known density function. Eq. (3) accounts for variations of the effects of the factors $X_{in}$, related to a specific injury severity level, in truck-involved crash probabilities for each lighting condition and area type model, where $\beta_i$ is determined using the density function $f(\beta_i|\phi)$. The mixed logit probabilities are calculated using weighted average for different values of $\beta_i$ across observations. Typically, some elements of $\beta_i$ are fixed and some are randomly distributed with specific statistical distribution. If the variance of $\phi$ is statistically significant, then the modeled injury severity levels vary with respect to $X$ across observations (Washington et al., 2003).

In this study, maximum likelihood estimation is performed through a simulation-based approach to overcome the computation complexity of estimating the parameters $\beta_i$ of the mixed logit models. The simulation procedure requires Halton draws. Compared to the purely random draws, Halton draws provide a more efficient distribution for numerical integration (Bhat, 2003; Halton, 1960). In addition to parameter estimation of the mixed logit models, marginal effects are estimated for the variables included in the model specifications. The marginal effects are computed as derivatives of the probability of injury severity level $i$ with respect to attribute $k$ in alternative $m$ (Greene, 2003).

$$\frac{\partial P_i}{\partial X_{km}} = [Q(i=m) - P_m]P_i\beta_k, \qquad i,m \in I \tag{4}$$

where the function $Q(i=m)$ equals 1 if $i$ equals $m$ and 0 otherwise. $P_i$ and $P_m$ denote the probability of injury severity level $i$ and $m$ ($i, m \in I$), respectively.

## 4. Data and empirical setting

Five years of crash records (2009 to 2013) involving trucks in the state of Ohio, provided by the Highway Safety Information System (HSIS), are used in this study. HSIS provides highway patrol reported data about crashes, and information about occupants, vehicles, and roadways involved in the crash.

The severity of the crashes is recorded as one of five injury levels. They are commonly defined using the KABCO injury scale: fatality (K), disabling injury (A), evident injury (B), possible injury (C), and no injury (O). Fatal injury includes crashes which result in death of occupant(s) within 30 days of crash. Disabling injury prevents the injured person from walking, driving or doing normal activities s/he was capable of performing before the injury. Evident injury includes crashes where injury is evident to observers at the crash location. Possible injury is one where occupant(s) complained of pain, but it diminishes rapidly from the time of evaluation at the crash location to the time of examination at the hospital. Lastly, no injury is where the reported crash does not result in any injury. The KABCO injury codes presented in the dataset are consolidated into three levels—major injury (KA), minor injury (B) and possible/no injury (CO). This approach is commonly used by researchers to ensure sufficient sample size for model estimation (e.g., Chen and Chen, 2011; Islam et al., 2014; Milton et al., 2008).

The effect of lighting condition on injury severity based on rural and urban roadways is the focus of this study. Hence, the analysis examined two area types: rural and urban, and three different lighting conditions: daylight, dark, and dark-lighted. To accomplish this, the dataset was first divided into rural and urban categories. Then for each category, the dataset was further divided into the three lighting conditions. The daylight dataset includes all of the crashes that occurred in the daylight period, except for those that occurred during dawn and dusk. The dark dataset includes crashes that occurred in dark condition without street lights, and the dark-lighted dataset includes crashes that occurred in dark condition with street lights. Based on the above classifications, six separate scenarios were considered: (1) rural daylight, (2) rural dark, (3) rural dark-lighted, (4) urban daylight, (5) urban dark, and (6) urban dark-lighted. Note that the crashes occurred during dawn and dusk were not considered in the analysis.

The final dataset consists of 41,461 observations. Each observation is a crash record that records the injury severity of the most severely injured occupant involved in the crash, along with occupant, vehicle, collision, roadway, and temporal and environmental characteristics. Hence, the dependent variable is the injury severity of the most severely injured occupant. There are 462 crashes involving major injury (1.1%), 1,705 crashes involving minor injury (4.1%), and 39,294 crashes involving possible/no injury (94.8%). Of these, 11,030 (26.6%) occurred during rural daylight condition, 4,429 (10.7%) occurred during rural dark condition, 822 (2.0%) occurred during rural dark-lighted condition, 20,122 (48.5%) occurred during urban daylight condition, 2,081 (5.0%) occurred during urban dark condition, and 2,977 (7.2%) occurred during urban dark-lighted condition.

**Table 2**
Descriptive statistics of variables by area type and lighting condition.

(a)

| Variables and description | Rural | | | Urban | | |
|---|---|---|---|---|---|---|
| | Daylight | Dark | Dark-lighted | Daylight | Dark | Dark-lighted |
| *Injury severity* | | | | | | |
| Major injury (1 if true; 0 otherwise) | 1.7% | 1.8% | 0.5% | 0.7% | 1.2% | 1.1% |
| Minor injury (1 if true; 0 otherwise) | 6.8% | 6.7% | 3.5% | 2.3% | 4.1% | 2.3% |
| Possible/no injury (1 if true; 0 otherwise) | 91.5% | 91.4% | 96.0% | 97.0% | 94.7% | 96.6% |
| | | | | | | |
| *Occupant characteristics* | | | | | | |
| Age (1 if age 35–45; 0 otherwise) | 24.6% | - | - | - | - | - |
| Age (1 if age 55–65; 0 otherwise) | - | - | - | 17.5% | - | 16.1% |
| Gender (1 if male; 0 otherwise) | 96.8% | - | 96.6% | - | - | 96.4% |
| Seating position (1 if seated at front; 0 otherwise) | 99.6% | - | - | - | - | 99.2% |
| Restraint use (1 if lap and/or shoulder belt used; 0 otherwise) | 95.6% | 96.9% | - | 93.8% | 96.2% | 93.8% |
| | | | | | | |
| *Vehicle characteristics* | | | | | | |
| Damage (1 if damaged; 0 otherwise) | 80.5% | - | - | - | 89.4% | 82.0% |
| Single-unit truck (1 if single-unit truck; 0 otherwise) | 26.6% | - | 10.3% | - | - | - |
| Truck trailer (1 if truck trailer; 0 otherwise) | - | 9.8% | - | 14.5% | - | - |
| Tractor semi-trailer (1 if tractor semi-trailer; 0 otherwise) | - | - | 76.8% | 46.8% | - | - |
| | | | | | | |
| *Collision characteristics* | | | | | | |
| Rear-end (1 if rear-end collision; 0 otherwise) | - | - | - | - | 9.1% | 15.1% |
| Sideswipe (1 if sideswipe collision—both meeting and passing; 0 otherwise) | 24.5% | 18.0% | - | 36.9% | - | 40.0% |
| Animal (1 if collision with an animal; 0 otherwise) | - | 30.1% | - | - | - | - |
| Object (1 if collision with roadside objects; 0 otherwise) | - | 22.0% | 23.4% | 10.5% | - | 15.8% |
| Motor vehicle in transport (1 if collision with motor vehicle in transport; 0 otherwise) | 48.6% | - | - | - | 43.5% | - |
| | | | | | | |
| *Roadway characteristics* | | | | | | |
| Surface type (1 if asphaltic concrete surface; 0 otherwise) | - | 95.8% | - | 93.3% | 95.4% | - |
| Curve (1 if in curve; 0 otherwise) | 12.8% | - | - | 10.1% | - | - |
| | | | | | | |
| *Temporal and environmental characteristics* | | | | | | |
| Weekday (1 if crash occurred on weekdays; 0 otherwise) | - | - | 84.3% | 90.9% | - | - |
| 12 AM to 4 AM (1 if crash occurred between 12 AM and 4 AM; 0 otherwise) | - | 27.8% | - | - | - | - |
| 8 AM to noon (1 if crash occurred between 8 AM and noon; 0 otherwise) | - | - | - | 34.9% | - | - |
| Noon to 4 PM (1 if crash occurred between noon and 4 PM; 0 otherwise) | - | - | - | 39.2% | - | - |
| Clear weather (1 if clear weather; 0 otherwise) | 82.2% | - | - | 83.9% | - | - |
| Adverse weather (1 if rain, snow, fog, and heavy-wind condition; 0 otherwise) | 17.3% | - | - | - | - | - |

(b)

| Variables and description | Mean | Standard deviation | Minimum | Maximum |
|---|---|---|---|---|
| *Roadway characteristics* | | | | |
| LogAADT (AADT varied between 110 and 185,730 veh/day) | | | | |
|     Rural daylight | 9.1 | 1.2 | 5.0 | 11.2 |
|     Urban daylight | 10.3 | 1.0 | 4.7 | 12.1 |
| Urban dark | 10.5 | 0.8 | 6.8 | 12.1 |
| Urban dark-lighted | 10.7 | 1.0 | 7.4 | 12.1 |
| Speed limit/10 (Speed limit varied between 20 and 70 mph) | | | | |
| Rural dark | 6.0 | 0.7 | 2.5 | 7 |
| Urban dark | 6.1 | 0.8 | 2.5 | 7 |
| Urban dark-lighted | 5.3 | 1.3 | 2.5 | 7 |
| Number of lanes (Number of lanes varied between 1 and 11) | | | | |
| Rural dark | 3.6 | 1.4 | 2 | 7 |
| Urban daylight | 4.5 | 1.7 | 1 | 11 |
| Urban dark-lighted | 5.0 | 1.8 | 2 | 11 |
| Surface width/10 (surface width varied between 15 and 144 ft) | | | | |
| Urban daylight | 5.7 | 2.0 | 1.5 | 14.4 |

Descriptive statistics of the variables used in the models are presented in Table 2. Part (a) of the table shows the frequency distribution of the indicator variables and part (b) shows the mean, standard deviation, minimum and maximum values of the continuous variables included in the models. For example, in case of urban daylight crashes, 90.9% of the crashes occurred during weekdays and 9.1% of the crashes occurred during weekend.

## 5. Model specification tests

The method often used to check the suitability of separate models over one aggregate model is to use likelihood ratio tests (Islam et al., 2014; Pahukula et al., 2015). In this study, once the six models were developed, a series of likelihood ratio tests were performed following the procedures articulated in Washington et al. (2003). Specifically, the tests were:

(i) the full model for all truck-involved crashes vs. the six separate models (rural daylight, rural dark, rural dark-lighted, urban daylight, urban dark, and urban dark-lighted);

(ii) the full model for all rural truck-involved crashes vs. the three separate models developed for rural locations (rural daylight, rural dark, rural dark-lighted);

(iii) the full model for all urban truck-involved crashes vs. the three separate models developed for urban locations (urban daylight, urban dark, and urban dark-lighted); and

(iv) all combinations of the three models within each area type (i.e., rural daylight vs. rural dark, rural dark vs. rural dark-lighted, rural daylight vs. rural dark-lighted, urban daylight vs. urban dark, urban dark vs. urban dark-lighted, and urban daylight vs. urban dark-lighted).

Likelihood ratio tests results are provided in Table 3. The first log-likelihood ratio test for transferability of coefficients from the full model to six separate lighting condition models is as follows.

$$LR_{Full} = -2\left[LL_{Full}(\beta^{Full}) - \sum_{j \in R \cup U} LL_j(\beta^j)\right] \tag{5}$$

where $LL_{Full}(\beta^{Full})$ is the log-likelihood at convergence of the full model (−7694), $LL_j(\beta^j)$ is the log-likelihood at convergence of subgroup $j$ using the same variables included in the full model, $R$ is the subgroups related to rural locations, and $U$ is the subgroups related to urban locations ($\sum_{j \in R \cup U} LL_j(\beta^j) =$ −7443). The $LR$ statistic ($LR = 251$) is $\chi^2$ distributed, with degrees of freedom ($df$) equal to the summation of the number of estimated parameters in all six models minus the number of estimated parameters in the full model. The null hypothesis here is that there is no difference in the parameter values between the full model and separate models. Chi-square statistic with 105 degrees of freedom resulted in a value greater than the critical value at the 99% confidence level ($\chi^2 = 141.62$), indicating that the models have statistically different model parameters.

The second and third log-likelihood ratio tests for transferability use the following equations.

$$LR_{Rural} = -2\left[LL_{Rural}(\beta^{Rural}) - \sum_{j \in R} LL_j(\beta^j)\right] \tag{6}$$

$$LR_{Urban} = -2\left[LL_{Urban}(\beta^{Urban}) - \sum_{j \in U} LL_j(\beta^j)\right] \tag{7}$$

where $LL_{Rural}(\beta^{Rural})$ is the log-likelihood at convergence of the full rural model, $LL_{Urban}(\beta^{Urban})$ is the log-likelihood at convergence of the full urban model, $LL_j(\beta^j)$ is the log-likelihood at convergence of subgroup $j$ ($j \in R$ is for rural locations and $j \in U$ is for urban locations). As presented in Table 3(a), the $LR$ for both rural and urban models are greater than the critical $\chi^2$ value at the 99% confidence level with their corresponding $df$; thus, separate models for both rural and urban locations are warranted.

The last log-likelihood test used to test the transferability of coefficients from the full rural and urban model to each corresponding lighting conditions model uses the following equation.

$$LR_{k_1 k_2} = -2\left[LL_{k_1 k_2}(\beta^{k_1 k_2}) - LL_{k_1}(\beta^{k_1})\right] \tag{8}$$

**Table 3**
Model specification tests.

(a)

| | Log-likelihood at convergence | | | | | | | Test statistic | $df$ | Critical value | Comment |
|---|---|---|---|---|---|---|---|---|---|---|---|
| | Full model | Rural | | | Urban | | | ($LR$) | | | |
| | | Daylight | Dark | Dark-lighted | Daylight | Dark | Dark-lighted | | | | |
| All data | −7694 | −2957 | −1208 | −110 | −2389 | −359 | −420 | 251 | 105 | 141.6 | $LR > \chi^2$ |
| All rural data | −4290 | −2886 | −1242 | −114 | | | | 92 | 40 | 63.6 | $LR > \chi^2$ |
| All urban data | −3232 | | | | −2401 | −379 | −426 | 90 | 30 | 50.8 | $LR > \chi^2$ |

(b)

| $k_1$ | $k_2$ | | | | | |
|---|---|---|---|---|---|---|
| | Rural | | | Urban | | |
| | Daylight | Dark | Dark-lighted | Daylight | Dark | Dark-lighted |
| Daylight | 0 | 915 ($df = 13$) | 790 ($df = 9$) | 0 | 352 ($df = 12$) | 66 ($df = 17$) |
| Dark | 156 ($df = 16$) | 0 | 93 ($df = 9$) | 58 ($df = 16$) | 0 | 84 ($df = 17$) |
| Dark-lighted | 50 ($df = 16$) | 61 ($df = 13$) | 0 | 55 ($df = 16$) | 46 ($df = 12$) | 0 |

where $LL_{k_1 k_2}(\beta^{k_1 k_2})$ is the log-likelihood at convergence of a model using the converged parameters from $k_2$'s model (using $k_2$'s data) on lighting condition $k_1$'s data and $LL_{k_1}(\beta^{k_1})$ is the log-likelihood at convergence of the model using lighting condition $k_1$'s data. The $LR$ statistic with $df$ equal to the number of estimated parameters in $\beta^{k_1 k_2}$ tests the hypothesis that the models have different parameters. Table 3(b) shows the results of these tests. All of these tests reject null hypothesis at the 99% confidence level.

The combination of all four types of likelihood tests yields a good assessment of the statistical differences among the three lighting conditions and two area types. Hence, it can be concluded that six separate models are statistically justified at the 99% confidence level.

## 6. Estimation results

Prior to estimating mixed logit models, the Hausman test (Hausman and McFadden, 1984) was conducted to determine if the multinomial logit model (MNL) would be appropriate; MNL models are not suitable when the independence of irrelevant alternatives (IIA) property is violated. The Hausman test results indicated that the MNL models are not appropriate.

Six separate mixed logit models were estimated for truck-involved crashes: rural daylight, rural dark, rural dark-lighted, urban daylight, urban dark, and urban dark-lighted. Each model predicts three levels of injury severity: major injury, minor injury, and possible/no injury. A simulation-based maximum likelihood method was utilized to estimate parameters $\beta_i$ for the mixed logit models. To estimate random parameters, the normal distribution was considered and 500 Halton draws were used. During the model development process, variables were retained in the specification if they have *t*-statistics corresponding to the 90% confidence level or higher on a two-tailed *t*-test. The random parameters were retained if their standard deviations have *t*-statistics corresponding to the 90% confidence level or higher. Model estimation results are presented in Tables 4 through 9.

The McFadden pseudo ratios ($\rho^2$) measure the improvement by the full models over the intercept models (i.e., constant only models). The $\rho^2$ values in Tables 4 through 9 indicate excellent overall improvement in model goodness-of-fit (between 0.72 and 0.87). A total of 14 parameters were found to be statistically significant as random parameters among the six estimated mixed logit models. All of these random parameters were shown to be significantly different from zero with at least 90% confidence. These random variables account for unobserved heterogeneity. Furthermore, inclusion of a random variable may reveal that one portion of the observations have a higher probability of a certain injury severity while another portion of the observations have a lower probability of that injury severity.

**Table 4**
Mixed logit model of truck-involved crashes injury severity for the daylight condition in rural location.

| Meaning of variable | Coefficient | t-statistic | p-value | Marginal effects | | |
| --- | --- | --- | --- | --- | --- | --- |
| | | | | Major injury | Minor injury | Possible/no injury |
| *Defined for major injury* | | | | | | |
| Seating position | −2.92 | −5.18 | 0.000 | −0.0970 | 0.0938 | 0.0032 |
| LogAADT (standard deviation of parameter distribution) | −0.68 (0.30) | −4.55 (4.31) | 0.000 (0.000) | −0.1039 | 0.0999 | 0.0040 |
| Adverse weather (standard deviation of parameter distribution) | −1.36 (2.56) | −1.89 (3.04) | 0.058 (0.002) | 0.0041 | −0.0040 | 0.0000 |
| *Defined for minor injury* | | | | | | |
| Gender | 0.45 | 1.73 | 0.084 | −0.0142 | 0.0206 | −0.0064 |
| Damage | −4.31 | −6.12 | 0.000 | 0.1370 | −0.1999 | 0.0630 |
| Single-unit truck | −0.50 | −4.36 | 0.000 | 0.0050 | −0.0076 | 0.0026 |
| Curve | −0.75 | −5.12 | 0.000 | 0.0054 | −0.0079 | 0.0026 |
| *Defined for possible/no injury* | | | | | | |
| Constant | −4.86 | −4.75 | 0.000 | | | |
| Age group (35–45) | −0.49 | −2.41 | 0.016 | 0.0001 | 0.0012 | −0.0013 |
| Restraint use | −2.38 | −13.7 | 0.000 | 0.0018 | 0.0255 | −0.0273 |
| Motor vehicle in transport | −0.46 | −2.67 | 0.008 | 0.0001 | 0.0023 | −0.0024 |
| LogAADT | −0.12 | −1.89 | 0.059 | 0.0011 | 0.0155 | −0.0166 |
| Sideswipe | −0.74 | −2.84 | 0.005 | 0.0001 | 0.0012 | −0.0012 |
| Clear weather | 0.61 | 2.48 | 0.013 | −0.0006 | −0.0079 | 0.0086 |
| *Model statistics* | | | | | | |
| Number of observations | 11,030 | | | | | |
| Restricted Log-likelihood (constant only) | −12,117.69 | | | | | |
| Log-likelihood at convergence | −3,084.30 | | | | | |
| McFadden pseudo R-squared ($\rho^2$) | 0.745 | | | | | |

Table 4 shows the estimation results of injury severity for the rural daylight condition model. Adverse weather variable was found to be normally distributed with mean −1.36 and standard deviation 2.56. These values indicate that for 29.8% of crashes under rural daylight condition occurring due to adverse weather, the probability of major injury was higher, and for the rest of the sample the probability of major injury was lower. Hence, for crashes due to adverse weather, the majority had a lower likelihood of major injury. The marginal effects of the variables included in the model are also presented in Table 4. They indicate the effects of one unit of change of one variable on each injury severity level. The interpretation of marginal effect is that if it is negative then there is a lower likelihood of incurring that injury severity level. Conversely, if the marginal effect is positive then there is a higher likelihood of incurring that injury severity level. For example, the gender variable has a positive value (0.0206) for minor injury. This means that if the occupant is male, then the probability of him experiencing a minor injury in a truck-involved crash in higher (2.06% higher than female). Note that the marginal effects of the gender variable for the major and possible/no injury levels would be negative. Specifically, if the occupant is a male, then his probability of experiencing a major injury is lower (1.42%) and experiencing a possible/no injury is also lower (0.64%). In this example, the increased likelihood of a minor injury is balanced out by the decreased likelihoods in major and possible/no injury. Note that in general, if the marginal effect is positive for one injury severity level, then it will be negative for the other two injury severity levels, and vice-versa.

Table 5 presents the estimation results for the rural dark condition model. For minor injury, the variable collision with an animal was found to be random and normally distributed with mean 2.78 and standard deviation 1.47. Given these estimates, for 2.9% of crashes due to collision with an animal under rural dark condition, the probability of minor injury was higher, and for rest of the observations the probability of minor injury was lower. This result implies that for crashes due to collision with an animal, the majority had a lower likelihood of being involved in a minor injury.

Table 6 shows the mixed logit model estimation results for crashes under rural dark-lighted condition. For major injury, the occupant being male was found to be random and normally distributed with mean −5.99 and standard deviation 3.65. With these parameters, 5.0% of the observations had a higher probability of being involved in a major injury while the rest of the observations had a lower probability of being involved in a major injury. This implies that for crashes where male occupants were involved, the majority had a lower likelihood of being a major injury. For minor injury, weekday was found to be random and normally distributed with mean 3.80 and standard deviation 3.13. Given these parameters, 11.2% of the crashes occurring on weekdays under rural dark-lighted condition had higher probability of minor injury and 88.8% of the crashes had lower probability of minor injury. This result implies that for crashes occurring during weekdays, the majority had a lower likelihood of being a minor injury.

**Table 5**
Mixed logit model of truck-involved crashes injury severity for the dark condition in rural location.

| Meaning of variable | Coefficient | t-statistic | p-value | Marginal effects | | |
|---|---|---|---|---|---|---|
| | | | | Major injury | Minor injury | Possible/no injury |
| *Defined for major injury* | | | | | | |
| Constant | −1.63 | −1.77 | 0.077 | | | |
| Truck trailer | −0.42 | −1.74 | 0.082 | −0.0019 | 0.0018 | 0.0001 |
| Restraint use | −1.27 | −4.77 | 0.000 | −0.0610 | 0.0591 | 0.0019 |
| Speed limit/10 | 0.61 | 3.78 | 0.001 | 0.2164 | −0.2098 | −0.0065 |
| Number of lanes (standard deviation of parameter distribution) | −0.44 (0.26) | −2.29 (2.02) | 0.021 (0.044) | −0.0491 | 0.0475 | 0.0016 |
| *Defined for minor injury* | | | | | | |
| Speed limit/10 | 0.42 | 7.71 | 0.000 | −0.1094 | 0.1496 | −0.0401 |
| 12 AM to 4 AM | −0.36 | −2.73 | 0.006 | 0.0056 | −0.0075 | 0.0019 |
| Animal (standard deviation of parameter distribution) | 2.78 (1.47) | 5.00 (2.54) | 0.000 (0.011) | −0.0034 | 0.0040 | −0.0007 |
| *Defined for possible/no injury* | | | | | | |
| Sideswipe | −0.99 | −2.39 | 0.017 | 0.0001 | 0.0014 | −0.0015 |
| Object | 0.55 | 2.33 | 0.020 | −0.0004 | −0.0048 | 0.0053 |
| Surface type | −1.34 | −4.15 | 0.000 | 0.0018 | 0.0197 | −0.0215 |
| *Model statistics* | | | | | | |
| Number of observations | 4,429 | | | | | |
| Restricted Log-likelihood (constant only) | −4,865.75 | | | | | |
| Log-likelihood at convergence | −1,359.13 | | | | | |
| McFadden pseudo R-squared ($\rho^2$) | 0.721 | | | | | |

**Table 6**
Mixed logit model of truck-involved crashes injury severity for the dark-lighted condition in rural location.

| Meaning of variable | Coefficient | t-statistic | p-value | Marginal effects | | |
|---|---|---|---|---|---|---|
| | | | | Major injury | Minor injury | Possible/no injury |
| *Defined for major injury* | | | | | | |
| Gender (standard deviation of parameter distribution) | −5.99 (3.65) | −2.27 (1.80) | 0.023 (0.072) | −0.0230 | 0.0215 | 0.0015 |
| Truck semi-trailer | −1.53 | −1.72 | 0.086 | −0.0186 | 0.0180 | 0.0005 |
| *Defined for minor injury* | | | | | | |
| Single-unit truck | 5.67 | 2.00 | 0.045 | −0.0017 | 0.0047 | −0.0029 |
| Object | −2.26 | −2.19 | 0.028 | 0.0113 | −0.0154 | 0.0041 |
| Weekday (standard deviation of parameter distribution) | 3.80 (3.13) | 1.98 (1.88) | 0.047 (0.060) | −0.0092 | 0.0032 | 0.0059 |
| *Defined for possible/no injury* | | | | | | |
| Constant | −6.63 | −4.02 | 0.000 | | | |
| Single-unit truck | 6.66 | 1.67 | 0.096 | −0.001 | −0.0063 | 0.0065 |
| *Model statistics* | | | | | | |
| Number of observations | 822 | | | | | |
| Restricted Log-likelihood (constant only) | −903.06 | | | | | |
| Log-likelihood at convergence | −126.29 | | | | | |
| McFadden pseudo R-squared ($\rho^2$) | 0.860 | | | | | |

**Table 7**
Mixed logit model of truck-involved crashes injury severity for the daylight condition in urban location.

| Meaning of variable | Coefficient | t-statistic | p-value | Marginal effects | | |
| --- | --- | --- | --- | --- | --- | --- |
| | | | | Major injury | Minor injury | Possible/no injury |
| *Defined for major injury* | | | | | | |
| Restraint use | −1.21 | −10.24 | 0.000 | −0.0170 | 0.0168 | 0.0002 |
| Truck trailer | −0.38 | −2.55 | 0.011 | −0.0015 | 0.0015 | 0.0000 |
| Sideswipe | −1.14 | −8.58 | 0.000 | −0.0026 | 0.0026 | 0.0000 |
| Clear weather | −0.71 | −6.85 | 0.000 | −0.0121 | 0.0119 | 0.0002 |
| Surface type | −0.95 | −7.84 | 0.000 | −0.0170 | 0.0168 | 0.0002 |
| Curve | 1.28 | 11.35 | 0.000 | 0.0035 | −0.0035 | 0.0000 |
| Noon to 4 PM | −0.27 | −2.59 | 0.009 | −0.0065 | 0.0065 | 0.0000 |
| *Defined for minor injury* | | | | | | |
| Age group (55–65) | 0.23 | 1.85 | 0.064 | −0.0008 | 0.0011 | −0.0002 |
| Truck semi-trailer | 0.23 | 2.44 | 0.015 | −0.0037 | 0.0046 | −0.0009 |
| 8 AM to noon (standard deviation of parameter distribution) | 1.72 (2.02) | 3.56 (5.47) | 0.001 (0.000) | −0.0014 | 0.0015 | −0.0001 |
| Object | −0.71 | −6.47 | 0.000 | 0.0052 | −0.0067 | 0.0014 |
| Weekday | 0.96 | 9.18 | 0.000 | −0.0147 | 0.0190 | −0.0043 |
| *Defined for possible/no injury* | | | | | | |
| LogAADT | −0.47 | −14.75 | 0.000 | 0.0013 | 0.0305 | −0.0318 |
| Number of lanes | 0.62 | 3.48 | 0.001 | −0.0008 | −0.0180 | 0.0187 |
| Surface width | −0.36 | −2.38 | 0.017 | 0.0005 | 0.0126 | −0.0131 |
| *Model statistics* | | | | | | |
| Number of observations | 20,122 | | | | | |
| Restricted Log-likelihood (constant only) | −22,106.27 | | | | | |
| Log-likelihood at convergence | −2,938.57 | | | | | |
| McFadden pseudo R-squared ($\rho^2$) | 0.867 | | | | | |

The model estimation results for crashes under urban daylight condition are presented in Table 7. For minor injury, the variable indicating crashes occurring between 8 AM and noon was found to be random and normally distributed with mean 1.72 and standard deviation 2.02. That means 19.7% of the crashes occurring between 8 AM and noon under rural dark-lighted condition had higher probability of minor injury and 80.3% of the crashes had lower probability of minor injury. Thus, for crashes occurring between 8 AM and noon, the majority had a lower likelihood of being a minor injury.

**Table 8**
Mixed logit model of truck-involved crashes injury severity for the dark condition in urban location.

| Meaning of variable | Coefficient | t-statistic | p-value | Marginal effects | | |
|---|---|---|---|---|---|---|
| | | | | Major injury | Minor injury | Possible/no injury |
| *Defined for major injury* | | | | | | |
| LogAADT (standard deviation of parameter distribution) | −2.31 (0.58) | −2.75 (2.16) | 0.006 (0.031) | −0.1912 | 0.1876 | 0.0037 |
| *Defined for minor injury* | | | | | | |
| Damage (standard deviation of parameter distribution) | −7.11 (0.84) | −2.18 (3.13) | 0.029 (0.002) | 0.0917 | −0.1455 | 0.0538 |
| Rear-end | −3.51 | −2.46 | 0.014 | 0.0069 | −0.0099 | 0.0030 |
| Speed limit/10 | −0.87 | −2.12 | 0.034 | 0.0675 | −0.1060 | 0.0385 |
| *Defined for possible/no injury* | | | | | | |
| Restraint use | −5.35 | −3.12 | 0.002 | 0.0008 | 0.0282 | −0.0289 |
| Motor vehicle in transport | −3.28 | −2.42 | 0.016 | 0.0002 | 0.0042 | −0.0044 |
| LogAADT (standard deviation of parameter distribution) | −1.06 (0.12) | −2.59 (3.55) | 0.009 (0.000) | 0.0023 | 0.0716 | −0.0739 |
| Surface type (standard deviation of parameter distribution) | −4.33 (2.61) | −2.44 (2.44) | 0.015 (0.015) | 0.0002 | 0.0045 | −0.0047 |
| *Model statistics* | | | | | | |
| Number of observations | 2,081 | | | | | |
| Restricted Log-likelihood (constant only) | −2,286.21 | | | | | |
| Log-likelihood at convergence | −421.43 | | | | | |
| McFadden pseudo R-squared ($\rho^2$) | 0.816 | | | | | |

Table 8 shows the model estimation results for crashes under urban dark condition. For minor injury, the variable indicating damage to vehicle was random and normally distributed with mean −7.11 and standard deviation 0.84. With these parameters, for 0.1% of the crashes under urban dark condition, the probability of minor injury was higher, and for rest of the crashes the probability of minor injury was lower. This implies that for crashes causing damage to vehicles, almost all of them had a lower likelihood of being a minor injury. For possible/no injury, the probability was higher for 4.9% of the crashes occurred on asphaltic concrete surface, and the probability was lower for rest of the crashes. The majority of the crashes occurred on asphaltic concrete surface had a lower likelihood of being a possible/no injury.

Table 9 presents the estimation results of mixed logit model for crashes under urban dark-lighted condition. For major injury, the occupant being male was random and normally distributed with mean 1.52 and standard deviation 0.24. These values indicate that for about 0.1% of the crashes involving male occupants, the probability of a major injury was higher, and for rest of the crashes the probability of a major injury was lower. Thus, for crashes involving male occupants, almost all of them had a lower likelihood of being a major injury. For possible/no injury, sideswipe collision was random and normally distributed with mean −6.71 and standard deviation 3.72. Given these values, for 3.6% of the crashes due to sideswipe collision under urban dark-lighted condition, the probability of possible/no injury was higher, and for the rest of the crashes the probability of possible/no injury was lower. This implies that for crashes due to sideswipe collision, the majority had a lower likelihood of being a possible/no injury.

**Table 9**
Mixed logit model of truck-involved crashes injury severity for the dark-lighted condition in urban location.

| Meaning of variable | Coefficient | $t$-statistic | $p$-value | Marginal effects | | |
|---|---|---|---|---|---|---|
| | | | | Major injury | Minor injury | Possible/no injury |
| *Defined for major injury* | | | | | | |
| Gender (standard deviation of parameter distribution) | 1.52 (0.24) | 1.66 (2.52) | 0.098 (0.002) | 0.0330 | −0.0324 | −0.0007 |
| Age group (55–65) | −1.74 | −3.02 | 0.003 | −0.0050 | 0.0048 | 0.0002 |
| LogAADT | −0.54 | −4.86 | 0.000 | −0.1285 | 0.1256 | 0.0029 |
| Speed limit/10 | 0.30 | 2.25 | 0.025 | 0.0363 | −0.0355 | −0.0008 |
| Damage | 1.53 | 2.60 | 0.009 | 0.0327 | −0.0320 | −0.0007 |
| *Defined for minor injury* | | | | | | |
| Gender | 1.94 | 3.89 | 0.000 | −0.0400 | 0.0553 | −0.0153 |
| Age group (55–65) | −1.45 | −3.15 | 0.002 | 0.0040 | −0.0082 | 0.0042 |
| Seating position | 1.16 | 2.27 | 0.023 | −0.0251 | 0.0348 | −0.0098 |
| Object (standard deviation of parameter distribution) | −0.82 (0.06) | −3.31 (3.49) | 0.001 (0.001) | 0.0052 | −0.0084 | 0.0032 |
| Number of lanes (standard deviation of parameter distribution) | −0.03 (0.16) | −0.29 (1.67) | 0.772 (0.095) | 0.0164 | −0.0221 | 0.0056 |
| *Defined for possible/no injury* | | | | | | |
| Restraint use | −2.61 | −5.74 | 0.000 | 0.0009 | 0.0152 | −0.0160 |
| Rear-end | −1.06 | −1.70 | 0.089 | 0.0000 | 0.0010 | −0.0010 |
| Sideswipe (standard deviation of parameter distribution) | −6.71 (3.72) | −0.56 (2.89) | 0.491 (0.004) | 0.0000 | −0.0015 | 0.0015 |
| *Model statistics* | | | | | | |
| Number of observations | 2,977 | | | | | |
| Restricted Log-likelihood (constant only) | −3,270.57 | | | | | |
| Log-likelihood at convergence | −466.72 | | | | | |
| McFadden pseudo R-squared ($\rho^2$) | 0.857 | | | | | |

**7. Discussion**

Separate models of injury severity levels by area types and lighting conditions provide valuable insights about contributing factors affecting the injury severity of truck-involved crashes. The model results suggest major differences in both the combination and the magnitude of impact of variables included in each model. Some variables are significant only in one lighting condition but not in other conditions. Similarly, some variables are found to be significant in one area type but not in other area type. These differences show that the different lighting conditions and area types do in fact have different contributing effects on injury severity in truck-involved crashes, further highlighting the importance of examining crashes based on lighting conditions on rural and urban roadways. Table 10 summarizes the effects of the statistically significant factors on injury severity by area types and lighting conditions.

*7.1. Occupant characteristics*

The difference in the effect of occupant age is worth noting. Occupants with age between 35 and 45 were found to be significant only at rural locations, while older occupants with age between 55 and 65 were found to be significant only at urban locations. During daylight, occupants with age between 35 and 45 were found to have increased probability of being involved in possible/no injury at rural locations. Occupants with age between 55 and 65 were found to have higher probability of minor injury during daylight. They were found to have lower probability of major and minor injury under dark-lighted conditions. This is perhaps due to the combined effects of being cautious while driving at night, having more driving experience, and accounting for longer reaction time. Male occupants were found to sustain major or minor injuries under rural daylight, rural dark-lighted, and urban dark-lighted conditions.

**Table 10**
Model comparisons.

| Variable | Rural | | | Urban | | |
|---|---|---|---|---|---|---|
| | Daylight | Dark | Dark-lighted | Daylight | Dark | Dark-lighted |
| *Occupant characteristics* | | | | | | |
| Age (35–45) | ↓ (poss/no) | | | | | |
| Age (55–65) | | | | ↑ (minor) | | ↓ (major, minor) |
| Gender | ↑ (minor) | | ↓ (major) | | | ↑ (major, minor) |
| Seating position | ↓ (major) | | | | | ↑ (minor) |
| Restraint use | ↓ (poss/no) | ↓ (major) | | ↓ (major) | ↓ (poss/no) | ↓ (poss/no) |
| | | | | | | |
| *Vehicle characteristics* | | | | | | |
| Damage | ↓ (minor) | | | | ↓ (minor) | ↑ (major) |
| Single-unit truck | ↓ (minor) | | ↑ (minor, poss/no) | | | |
| Truck trailer | | ↓ (major) | | ↓ (major) | | |
| Tractor semi-trailer | | | ↓ (major) | ↑ (minor) | | |
| | | | | | | |
| *Collision characteristics* | | | | | | |
| Rear-end | | | | | ↓ (minor) | ↓ (poss/no) |
| Sideswipe | ↓ (poss/no) | ↓ (poss/no) | | ↓ (major) | | ↓ (poss/no) |
| Animal | | ↑ (minor) | | | | |
| Object | | ↑ (poss/no) | ↓ (minor) | ↓ (minor) | | ↓ (minor) |
| Motor vehicle in transport | ↓ (poss/no) | | | | ↓ (poss/no) | |
| | | | | | | |
| *Roadway characteristics* | | | | | | |
| LogAADT | ↓ (major, poss/no) | | | ↓ (poss/no) | ↓ (major, poss/no) | ↓ (major) |
| Speed limit/10 | | ↑ (major, minor) | | | ↓ (minor) | ↑ (major) |
| No. of lanes | | ↓ (major) | | ↑ (poss/no) | | ↓ (minor) |
| Surface type | | ↓ (poss/no) | | ↓ (major) | ↓ (poss/no) | |
| Curve | ↓ (minor) | | | ↑ (major) | | |
| Surface width/10 | | | | ↓ (poss/no) | | |
| | | | | | | |
| *Temporal and environmental characteristics* | | | | | | |
| Weekday | | | ↑ (minor) | ↑ (minor) | | |
| 12 AM to 4 AM | | ↓ (minor) | | | | |
| 8 AM to noon | | | | ↑ (minor) | | |
| Noon to 4 PM | | | | ↓ (major) | | |
| Clear weather | ↑ (poss/no) | | | ↓ (major) | | |
| Adverse weather | ↓ (major) | | | | | |

↑ indicates increase in the probability of an injury severity level; ↓ indicates decrease in the probability of an injury severity level; poss/no represents possible/no injury severity level.

The occupant seated at the front of the vehicle was associated with major injury under rural day light and minor injury under urban dark-lighted conditions. The use of lap and/or shoulder belt was found to decrease the likelihood of major injury under rural dark conditions. In contrast, it was found to decrease the likelihood of possible/no injury under rural daylight conditions. A possible reason for this is that crashes occurring at rural locations during nighttime are typically severe, which are likely to cause major injury, but the use of restraint reduces the severity. Under both urban dark and dark-lighted conditions, the use of lap and/or shoulder belt was negatively associated with possible/no injury.

*7.2. Vehicle characteristics*

Regarding vehicle types, single-unit truck was found to decrease minor injury during daylight and it was found to increase minor and possible/no injury under dark-lighted conditions at rural locations. Truck trailer was found to decrease major injury for both rural dark and urban daylight conditions. Lastly, tractor semi-trailers were found to decrease major injury for rural dark-lighted conditions, and they were found to increase minor injury for urban daylight conditions. It is evident that the crashes involving multiple unit trucks (i.e., truck trailer, tractor semi-trailer) are more severe during nighttime at rural locations. This is likely because multiple unit trucks are heavier (typically weighing much more than 10,000 lbs); thus, a higher probability of severe injury for the occupants.

*7.3. Collision characteristics*

Rear-end collision was found to decrease the probability of minor injury under urban dark conditions and possible/no injury under urban dark-lighted conditions. Sideswipe collision was associated with less severe injury at rural locations. Interestingly, sideswipe collision was associated with major injury under urban daylight conditions. This could be due to the fact that during the day time urban roadways carry high volume of traffic, which increases the probability of sideswipe collision. Animal involved crashes were found to be significant only under rural dark conditions. When a vehicle hits an animal, the probability of an occupant being involved in minor injury increases. Hitting objects was found to decrease the probability of minor injury under rural dark-lighted, urban daylight, and urban dark-lighted conditions. Lastly, collision with other motor vehicles in transport was negatively associated with possible/no injury under rural daylight and urban dark conditions.

*7.4. Roadway characteristics*

The variable LogAADT was found to be significant only under daylight conditions at rural locations. It was negatively associated with both major and possible/no injury, which means increased traffic at rural locations will decrease the probability of major and possible/no injury during day time. One possible explanation could be the fact that when traffic volume increases drivers will become more cautious, resulting in lower probability of major injury. Furthermore, high traffic volume may increase the probability of less severe injury. At urban locations, as the traffic volume increased the probability of possible/no injury decreased during day time, while the probability of both major and possible/no injury decreased during nighttime. One possible explanation could be the fact that drivers are more cautious while driving at night. Speed limit was positively associated with both major and minor injury under rural dark conditions. This is perhaps because of the higher impact speed in a collision. At urban locations, as the speed increased the probability of minor injury decreased under dark conditions and the probability of major injury increased under dark-lighted conditions.

As the number of lanes increased the probability of major injury was found to decrease under rural dark conditions. A possible reason for this is that more lanes provide drivers with more opportunities to avoid last minute collisions by changing lanes. Under urban daylight conditions, as the number of lanes increased the probability of possible/no injury was found to increase. Under urban dark-lighted conditions, as the number of lanes increased the probability of minor injury was found to decrease.

Asphaltic concrete surface was found to decrease the probability of possible/no injury under both rural and urban dark conditions. Furthermore, it was found to decrease the probability of major injury under urban daylight conditions. One important finding from these facts is that asphaltic concrete surface could reduce the likelihood of severe injury crashes during nighttime. Curved highways were found to decrease the probability of minor injury under rural daylight conditions. Under urban daylight conditions, curved roadways were found to increase the probability of major injury. This is perhaps because of the combined effects of severe collisions due to curved roadways and high traffic volume on urban roadways.

*7.5. Temporal and environmental characteristics*

The probability of minor injury increased for the crashes occurring during weekdays under rural dark-lighted and urban daylight conditions. This may be because urban roadways carry high traffic volume during weekdays. Clear weather was found to increase the probability of possible/no injury under rural daylight condition and decrease the probability of major injury under urban daylight condition. Another important finding from the clear weather variable is that dark and dark-lighted conditions were found not to be significant for both area types. Adverse weather condition variable was found to be significant only under rural daylight conditions. The probability of major injury decreased under adverse weather condition. One possible explanation could be that traffic tends to go slower in adverse weather conditions.

## 8. Conclusions

This study employed mixed logit (random parameters logit) modeling framework to investigate lighting condition and area type differences in the injury severity of crashes involving trucks. Using the data from the HSIS for the state of Ohio, separate models for two area types and three lighting conditions were developed: rural daylight, rural dark, rural dark-lighted, urban daylight, urban dark, and urban dark-lighted. A series of log-likelihood ratio tests were conducted to validate that these six separate models by lighting conditions and area types are warranted. The model estimation results demonstrated the necessity of using a disaggregate approach to analyze truck-involved crashes.

The model results suggest major differences in both the combination and the magnitude of impact of variables included in each model. Some variables are significant only in one lighting condition but not in other conditions. Similarly, some variables are found to be significant in one area type but not in other area type. Key differences include age and gender of occupant, types of trucks, speed, AADT, curve roadways, and adverse weather. For example, it was found that increasing speed causes an increase in both major and minor injury for rural dark condition, but it causes a decrease in minor injury for urban dark condition.

Separate injury severity models based on lighting conditions and area types for truck-involved crashes has yielded some new information not present in the exiting literature. However, similar to previous studies on safety analyses, this study also has some limitations which should be taken into account before applying its findings. One is that the crash data came from a single U.S. state, and second, the factors investigated were limited to those available in the HSIS database. The findings would be more generalizable if the dataset had crashes from multiple states and if it could be linked to other databases to provide additional information about the truck-involved crash injury severity under different lighting conditions in rural and urban roadways. For instance, the information related to the movements of truck just before crash occurrence (such as turning left, turning right, skidding and merging), defects related to truck (such as brakes defect, tires defect and cargo defect), etc. could be considered.

**Acknowledgment**

# References

Abdel-Aty, M., 2003. Analysis of Driver Injury Severity Levels at Multiple Locations using Ordered Probit Models. J. Safety Res. 34, 597–603. doi:10.1016/j.jsr.2003.05.009

Abramson, H., 2015. The Trucks Are Killing Us. New York Times. Accessed from: www.nytimes.com/ 2015/08/22/ opinion/the-trucks-are-killing-us.html

Anastasopoulos, P.C., Mannering, F.L., 2011. An Empirical Assessment of Fixed and Random Parameter Logit Models Using Crash- and Non-Crash-Specific Injury Data. Accid. Anal. Prev. 43, 1140–1147. doi:10.1016/ j.aap.2010.12.024

Bhat, C.R., 2003. Simulation Estimation of Mixed Discrete Choice Models Using Randomized and Scrambled Halton Sequences. Transp. Res. Part B Methodol. 37, 837–855. doi:10.1016/S0191-2615(02)00090-5

Bureau of Transportation Statistics, 2015. Freight Facts and Figures. Accessed from: www.rita.dot.gov/bts/sites/ rita.dot.gov.bts/files/FF&F_complete.pdf

Chang, L.-Y., Mannering, F., 1999. Analysis of Injury Severity and Vehicle Occupancy in Truck- and Non-Truck-Involved Accidents. Accid. Anal. Prev. 31, 579–592. doi:10.1016/S0001-4575(99)00014-7

Chen, F., Chen, S., 2011. Injury Severities of Truck Drivers in Single- and Multi-Vehicle Accidents on Rural Highways. Accid. Anal. Prev. 43, 1677–1688. doi:10.1016/j.aap.2011.03.026

Dong, C., Richards, S.H., Huang, B., Jiang, X., 2015. Identifying the Factors Contributing to the Severity of Truck-Involved Crashes. Int. J. Inj. Contr. Saf. Promot. 22, 116–126. doi:10.1080/ 17457300.2013.844713

Duncan, C.S., Khattak, A.J., Council, F.M., 1998. Applying the Ordered Probit Model to Injury Severity in Truck–Passenger Car Rear-End Collisions. Transp. Res. Rec. J. Transp. Res. Board 1635, 63–71. doi:10.3141/ 1635-09

Greene, W., 2003. Econometric Analysis, 5th ed. Prentice Hall, Upper Saddle River, New Jersey.

Halton, J.H., 1960. On the Efficiency of Certain Quasi-Random Sequences of Points in Evaluating Multi-Dimensional Integrals. Numer. Math. 2, 84–90. doi:10.1007/BF01386213

Hausman, J., McFadden, D., 1984. Specification Tests for the Multinomial Logit Model. Econometrica 52, 1219–1240. doi:10.2307/1910997

Islam, M., Hernandez, S., 2013a. Large Truck–Involved Crashes: Exploratory Injury Severity Analysis. J. Transp. Eng. 139, 596–604. doi:10.1061/(ASCE)TE.1943-5436.0000539

Islam, M., Hernandez, S., 2013b. Modeling Injury Outcomes of Crashes Involving Heavy Vehicles on Texas Highways. Transp. Res. Rec. J. Transp. Res. Board 2388, 28–36. doi:10.3141/2388-05

Islam, S., Jones, S.L., Dye, D., 2014. Comprehensive Analysis of Single- and Multi-Vehicle Large Truck At-Fault Crashes on Rural and Urban Roadways in Alabama. Accid. Anal. Prev. 67, 148–158. doi:10.1016/j.aap.2014.02.014

Khattak, A.J., Schneider, R.J., Targa, F., 2003. Risk Factors in Large Truck Rollovers and Injury Severity: Analysis of Single-Vehicle Collisions, in: TRB Annual Meeting CD-ROM. Transportation Research Board of the National Academies.

Khorashadi, A., Niemeier, D., Shankar, V., Mannering, F., 2005. Differences in Rural and Urban Driver-Injury Severities in Accidents Involving Large-Trucks: An Exploratory Analysis. Accid. Anal. Prev. 37, 910–921. doi:10.1016/j.aap.2005.04.009

Kim, J.K., Ulfarsson, G.F., Kim, S., Shankar, V.N., 2013. Driver-Injury Severity in Single-Vehicle Crashes in California: A Mixed Logit Analysis of Heterogeneity due to Age and Gender. Accid. Anal. Prev. 50, 1073–1081. doi:10.1016/j.aap.2012.08.011

Lemp, J.D., Kockelman, K.M., Unnikrishnan, A., 2011. Analysis of Large Truck Crash Severity Using Heteroskedastic Ordered Probit Models. Accid. Anal. Prev. 43, 370–380. doi:10.1016/j.aap. 2010.09.006

Lyman, S., Braver, E.R., 2003. Occupant Deaths in Large Truck Crashes in the United States: 25 Years of Experience. Accid. Anal. Prev. 35, 731–739. doi:10.1016/S0001-4575(02)00053-2

Milton, J.C., Shankar, V.N., Mannering, F.L., 2008. Highway Accident Severities and the Mixed Logit Model: An Exploratory Empirical Analysis. Accid. Anal. Prev. 40, 260–266. doi:10.1016/ j.aap.2007.06.006

NHTSA, 2015. Traffic Safety Facts 2013. Accessed from: http://www-nrd.nhtsa.dot.gov/Pubs/812139.pdf

Osman, M., Paleti, R., Mishra, S., Golias, M.M., 2016. Analysis of Injury Severity of Large Truck Crashes in Work Zones. Accid. Anal. Prev. 97, 261–273. doi: 10.1016/j.aap.2016.10.020

Pahukula, J., Hernandez, S., Unnikrishnan, A., 2015. A Time of Day Analysis of Crashes Involving Large Trucks in Urban Areas. Accid. Anal. Prev. 75, 155–163. doi:10.1016/j.aap.2014.11.021

Savolainen, P.T., Mannering, F.L., Lord, D., Quddus, M.A., 2011. The Statistical Analysis of Highway Crash-Injury Severities: A Review and Assessment of Methodological Alternatives. Accid. Anal. Prev. 43, 1666–1676. doi:10.1016/j.aap.2011.03.02527

Train, K., 2009. Discrete Choice Methods with Simulation, 2nd ed. Cambridge University Press, Cambridge, UK.

Washington, S.P., Karlaftis, M.G., Mannering, F.L., 2003. Statistical and Econometric Methods for Transportation Data Analysis, 1st ed. Chapman & Hall/CRC, Boca Raton.

Zaloshnja, E., Miller, T.R., 2007. Unit Costs of Medium and Heavy Truck Crashes. Report: FMCSA-RRA-07-034, Washington, DC. Accessed from: mcsac.fmcsa.dot.gov/documents/Dec09/ UnitCostsTruckCrashes2007.pdf

Zhu, X., Srinivasan, S., 2011a. A Comprehensive Analysis of Factors Influencing the Injury Severity of Large-Truck Crashes. Accid. Anal. Prev. 43, 49–57. doi:10.1016/j.aap.2010.07.007

Zhu, X., Srinivasan, S., 2011b. Modeling Occupant-Level Injury Severity: An Application to Large-Truck Crashes. Accid. Anal. Prev. 43, 1427–1437. doi:10.1016/j.aap.2011.02.021